\RequirePackage{fix-cm}
\documentclass[smallextended]{svjour3}       
\smartqed  
\usepackage{graphicx}
\usepackage{amssymb}
\usepackage{amsmath}
\newcommand{\be}{\begin{equation}}
\newcommand{\ee}[1]{\label{#1}\end{equation}}
\newcommand{\bem}{\begin{eqnarray}}
\newcommand{\eem}[1]{\label{#1}\end{eqnarray}}

\begin{document}

\title{Breakdown of zero-energy quantum Hall state in graphene in the light of current fluctuations and shot noise}


\titlerunning{Breakdown noise of $\nu = 0$ in suspended graphene Corbino disk}        

\author{Antti Laitinen \and Manohar Kumar \and Teemu Elo \and Ying Liu \and T. S. Abhilash \and Pertti J. Hakonen}


\institute{	Antti Laitinen \and Manohar Kumar \and Teemu Elo \and Ying Liu \and T. S. Abhilash \and Pertti J. Hakonen \at
              Low Temperature Laboratory, Department of Applied Physics, Aalto University, Espoo, Finland \\
              \email{antti.laitinen@aalto.fi}   \\
              \and
              Ying Liu \at
               Laboratory of Science and Technology on Integrated Logistics Support, College of Mechatronics and Automation, National University of Defense Technology, Changsha 410073, China\\
              \and
              Manohar Kumar \at
              \emph{Present address:}
              Laboratoire Pierre Aigrain - \'Ecole Normale sup\'erieure, 24 rue Lhomond, 75231 Paris Cedex 05, France \\
}


\date{Received: date / Accepted: date}

\maketitle

\begin{abstract}
We have investigated the cross-over  from Zener tunneling of single charge carriers to avalanche type of bunched electron transport in a suspended graphene Corbino disk in the zeroth Landau level. At low bias, we find a tunneling current that follows the gyrotropic Zener tunneling behavior. At larger bias, we find avalanche type of transport that sets in at a smaller current the larger the magnetic field is. The low-frequency noise indicates strong bunching of the electrons in the avalanches. On the basis of the measured low-frequency switching noise power, we deduce the characteristic switching rates of the avalanche sequence. The simultaneous microwave shot noise measurement also reveals  intrinsic correlations within the avalanche pulses and indicate decrease of correlations with increasing bias.
\end{abstract}

\section{Introduction}
\label{intro}
Breaking of the quantum Hall state in a 2D electron gas (2-DEG) is an intriguing topic as it mixes issues of edge and bulk transport. Several different models for quantum Hall state breakdown mechanisms have been proposed and they can be grouped into two main categories: bootstrap electron heating (BSEH) and quasi-elastic inter-Landau level scattering (QUILLS). There are experimental results supporting both of these views and no consensus has been reached yet. While the BSEH is usually associated with the edge states, the QUILLS mechanism is founded on the temperature dependence of the conductivity $\sigma_{xx}$ in the bulk of the 2DEG. Providently, in the Corbino sample geometry there are no continuous edge states across the sample, which makes this geometry particularly suitable for testing theories dependent on the bulk properties. Therefore Corbino geometry, implemented using an ultimate 2D material such as monolayer graphene, forms an excellent platform for studying the bias-induced breakup of the quantum Hall state.

The present paper extends the work of Ref. \cite{Laitinen2017} and  concentrates on the cross-over from the gyrotropic Zener tunneling (inter-LL Zener tunneling) to avalanche type of transport, and finally to nearly ohmic behavior. In our IV characteristics, the Zener tunneling regime ends in a sharp increase of current with the bias voltage, which is quite well in line with the expectations from the BSEH theories. We address this cross over regime using low frequency current fluctuations (frequency $f\leq 10$ Hz) and shot noise $S_I$ experiments (over microwave band $f=650-900$ MHz) in order to obtain further information on the charge transfer processes underlying these phenomena. Our results indicate strong low-frequency noise in the rapidly growing, steep sections of the IV curves, which we interpret as bunching of electrons in the avalanche generation of the BSEH regime. This low-frequency noise grows as average current squared $\langle I \rangle^2$, which is typical for switching type of noise, as well as for resistance fluctuations with uniform energy spectrum. In our region of interest, the inevitable contact resistance fluctuations can be neglected, and we analyze the observed low frequency Lorentzian noise spectrum as a switching process. This analysis allows us to determine characteristic values for the basic transition rates in the switching process. At microwave frequencies, on the other hand, the noise is close to regular shot noise, although we find predominantly levels below the full Poissonian noise. The amount of shot noise at microwave frequencies is dependent on the correlations among the charge carriers: with full temporal correlations the zero-frequency shot noise vanishes \cite{Blanter2000}. Our results indicate that close to the onset of the avalanche regime there are substantial correlations between the carriers, whereas approximately Poissonian noise is reached when the avalanche pulses start to overlap each other.

\subsection{Inter-Landau level Zener tunneling}
The zero-energy Landau level (see, \textit{e.g}. Ref. \cite{ezawa}) in suspended graphene provides a well-isolated setting for investigations of the breakdown of the quantum Hall state at magnetic fields $B$ around 1 T and even below. At small $B$, the energy scales in this unique quantum Hall state can be kept small, which allows sensitive quasi-elastic studies of the breakdown at mK temperatures. At higher temperatures, however, inelastic processes such as phonon-assisted scattering processes may start to dominate the breakdown. The zero-energy level is four-fold degenerate with respect to the two values of true electron spin and two values of pseudospin that describes the distribution of electrons between two valleys in the graphene Brillouin zone. The degeneracy, however, is lifted by Coulomb interactions and the original four-fold degenerate zero-energy states split into two states with a gap between them.  According to theoretical considerations  \cite{ezawa}, the insulator state was expected to be ferromagnetic, although the experiments of Refs. \cite{Giesbers2009} and \cite{young} did not confirm this. Independently from the character of the gap, the zero-energy Landau state is an insulator with the gap $\Delta$ of the order of the characteristic Coulomb energy equal to $e^2/\epsilon \ell_B$ in graphene. Here $\epsilon$ is the dielectric constant, $\ell_B =\sqrt{\Phi_0/2\pi B}$ is the magnetic length, $B$ is the magnetic field, and $\Phi_0=h/e$ is the single-electron flux quantum.

Any insulator can be driven to a conducting state using a high voltage $V>V_{cr}$, where $V_{cr}$ denotes the threshold for dielectric breakdown.  At bias voltages $V \gg V_{cr}$, the $IV$ curve becomes nearly linear (ohmic regime). At $V \ll V_{cr}$, on the other hand, the conductance is strongly suppressed and distinctly nonlinear. At low temperatures, an exponentially small current $I$ emerges due to Zener tunneling between the two bands \cite{Zener1934}, creating an electron in the empty upper band and a hole in the full lower band. \cite{ziman} Recently, we demonstrated \cite{Laitinen2017} that the regular equation for the Zener tunneling current  $ I\propto e ^{-V_Z/ V }$, with $V_Z \propto \Delta^{3/2} $ has to be replaced by a gyrotropic law given by
\be
  I\propto e ^{-(V_Z/V)^2 },
     \ee{vor}
where $V_Z=\frac{eBd}{\sqrt{8\pi}\epsilon\Phi_0}$. The formula follows from the gyrotropic Zener tunneling theory developed for the zero-energy Landau level of graphene in a strong magnetic field \cite{Laitinen2017}.  This behavior provides evidence that the quantum tunneling processes here are governed - instead of the particle mass - by the gyrotropic force on a particle. Such a behavior is similar to the motion of quantized vortices in superfluids, for example in the nucleation of vortices at a plane boundary \cite{Vol72,Son73}, with the Magnus force (analog of the Lorentz force on an electron) balancing the external force \cite{EBS}. The Zener tunneling processes in the quantum Hall regime of a 2-DEG  are central in the quasi-elastic inter-Landau level scattering (QUILLS) \cite{Heinonen1984,Eaves1984} and in the magnetoresistance oscillations in the ohmic regime \cite{Zener,Bykov2012}. The gyrotropic theory, however, is distinct from them in its implications for the IV-characteristics.  In small constrictions QUILLS type of behavior has been observed at one-micron dimensions \cite{Bliek1996,Makarovsky2002}, i.e. at similar length scales where we observe the gyrotropic behavior in our experiments.

In the studies of the quantum Hall state breakdown, the Corbino geometry has a distinct advantage over the Hall bar geometry, because no edge states persist and the breakdown is dominated by transport processes in the bulk. The Corbino geometry has been used for numerous studies of phenomena related to transitions between Landau levels in 2D electron gases, most notably in the works of microwave-induced resistance oscillations and zero-resistance states \cite{Yang2003}, as well as phonon-induced resistance oscillations \cite{Liu2014}. It has also been employed to investigate standard-type Zener tunneling between different Landau levels \cite{Goran2013} and the bootstrap electron heating (BSEH) model \cite{Komiyama1985}, which has experimentally been found to be a reasonable route to breakdown of the quantum Hall effect (QHE) \cite{Ebert1983,Cage1983,Nachtwei1999,Komiyama2000a}. Recently, the Corbino geometry has also been employed for measuring current fluctuations in order to investigate the nature of bulk transport phenomena at the onset of the integer QHE breakup. In their low-frequency current noise experiments on the first Landau level of a 2-DEG heterostructure, Kobayashi and coworkers \cite{Chida2014,Hata2016} found strong bunching of electrons at the onset of the breakdown of the quantum Hall state, supporting the BSEH view.

\subsection{Avalanche type of transport} \label{AVtheory}

\begin{figure}[htb]
\includegraphics[width=0.8\textwidth]{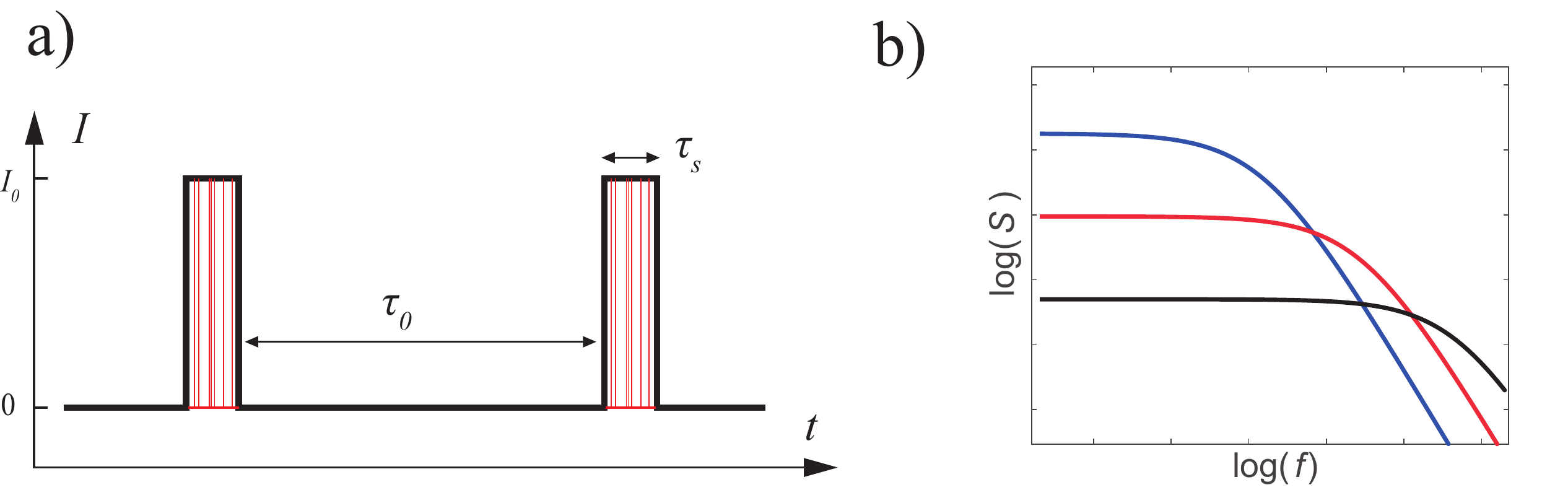}
\centering
\caption{a) Model of low-frequency avalanche current noise derived from a switching sequence of rectangular current pulses with the nominal height of $I_0$ having asymmetric upward and downward transition rates $1/\tau_0$ and $1/\tau_s$, respectively. This corresponds to pulses with mean duration $\tau_s$ separated by a waiting time $\tau_0$ on average. b) Schematic illustration of Eq. \ref{limit1} at a few different lengths of the avalanche pulse $\tau_s$ (inverse of transition rate): the chosen specific transition rates result in a spectrum which is  flat at low frequencies and rolls off linearly on the log-log scale of the illustration.}
\label{fig:switchingNoise}
\end{figure}

Avalanche type of current transport can be considered as one type of switching transport. Such transport is illustrated in Fig.  \ref{fig:switchingNoise}a, where the current consists of a random sequence of two current levels 0 and $I_0$. The average switching rate upwards is denoted by $1/\tau_0$ while the rate downwards is $1/\tau_s$, which means that the average duration of avalanche pulses is $\tau_s$. Following e.g. Ref. \cite{Kogan}, one can derive for the  avalanche type switching noise spectral  density $S_I^{AV}$ the equation:
\begin{equation}
S_I^{AV}(\omega) = 4\frac{(\tau_0 \tau_s)^2}{(\tau_0 + \tau_s)^3}I_0^2\frac{1}{1+\omega^2\left( \frac{\tau_0 \tau_s}{\tau_0 + \tau_s} \right)^2} .
\end{equation}
We are interested in the cross-over from Zener tunneling events to avalanche type of transport. Initially at small currents, the avalanche events are rare and we can make the approximation $\tau_0 \gg \tau_s$. Using this approximation and recognizing that $\left\langle I\right\rangle = \left(\frac{\tau_s}{\tau_0}\right) I_0$, we obtain the spectrum:
\begin{equation} \label{limit1}
S_I^{AV}(\omega) = 4\tau_0 \left\langle I\right\rangle^2 \frac{1}{1+\omega^2\tau_s^2},
\end{equation}
which is illustrated schematically in Fig.  \ref{fig:switchingNoise}b. The length of the avalanche pulse determines the corner frequency $\omega_c \propto 1/\tau_s$ in this regime and this possible variation in spectral density is indicated in Fig.  \ref{fig:switchingNoise}b at three values of $\omega_c$.

The time trace of Fig. 1 assumes that there is only  one scenario for producing the avalanche sequence of charge carriers. In reality, there might be several parallel scenarios, which would then lead to a sum of Lorentzian spectra, each with a different weighting factor. Such a scenario could eventually lead to $1/f$ type of noise spectrum.

The above low-frequency model neglects fluctuations within the avalanche pulse. The current during the avalanche pulse will consist of individual charge carrier events, the spacing of which may vary in time. These fluctuations lead to wide-band shot noise which becomes important at frequencies $\omega \gg 1/\tau_s$ where the Lorentzian fluctuator spectrum of Eq. \ref{limit1} has already decreased below the shot noise level. The shot noise intrinsic to the avalanche pulse will depend on the correlations between charge carriers in the pulse: if the pulse is fully correlated, there is no shot noise \footnote{However, there will be a peak in the noise power spectrum at the frequency corresponding to the inverse of the arrival period of the correlated charge carriers.}, while without any correlation there should be independent carrier emissions with a full Poissonian shot noise like in a tunnel junction. Hence, the Fano factor for the current during the avalanche pulse is expected to vary between $ F = 0 - 1$. Using the average current $\langle I \rangle=I_0\frac{\tau_s}{\tau_s+\tau_0}$, the shot noise spectral density arising from the current pulses can be written as:
\begin{equation}\label{noise_shot_avalanche}
  S_{sh}^{AV}(\omega)=F 2e\langle I \rangle.
\end{equation}
A further possibility would be to have an effective tunneling charge which is different from one electron. Here we assume this possibility to be included in to the value of the Fano factor.

\section{Experimental techniques}
Our measurements down to 10 mK were performed in a BlueFors LD-400 dilution refrigerator. The cryostat was equipped with a DC/audio-frequency measurement circuitry for IV characteristics and low-frequency noise. The bias leads were connected via microwave bias-T components to the sample, which facilitated simultaneous microwave noise measurements along a separate 50\,$\Omega$ output channel. The low frequency measurement leads were twisted pair phosphor-bronze wires supplemented by three stage $RC$ filters with a nominal cut-off given by $R=150$ $\Omega$ and $C=10$ nF. However, due the high impedance of the measured quantum Hall devices the actual cutoff is dependent on the resistance of the sample. For magneto-conductance scans, see Fig. \ref{fig:basics}, we used an AC peak-to-peak current excitation of $0.1$ nA at $f=3.333$ Hz.

\subsection{Samples and their characterization}
\begin{figure}[htb]
\includegraphics[width=0.6\textwidth]{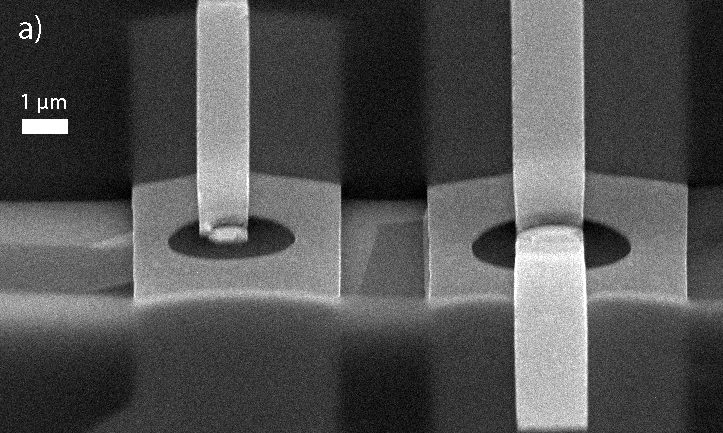}
\includegraphics[width=1.3\textwidth]{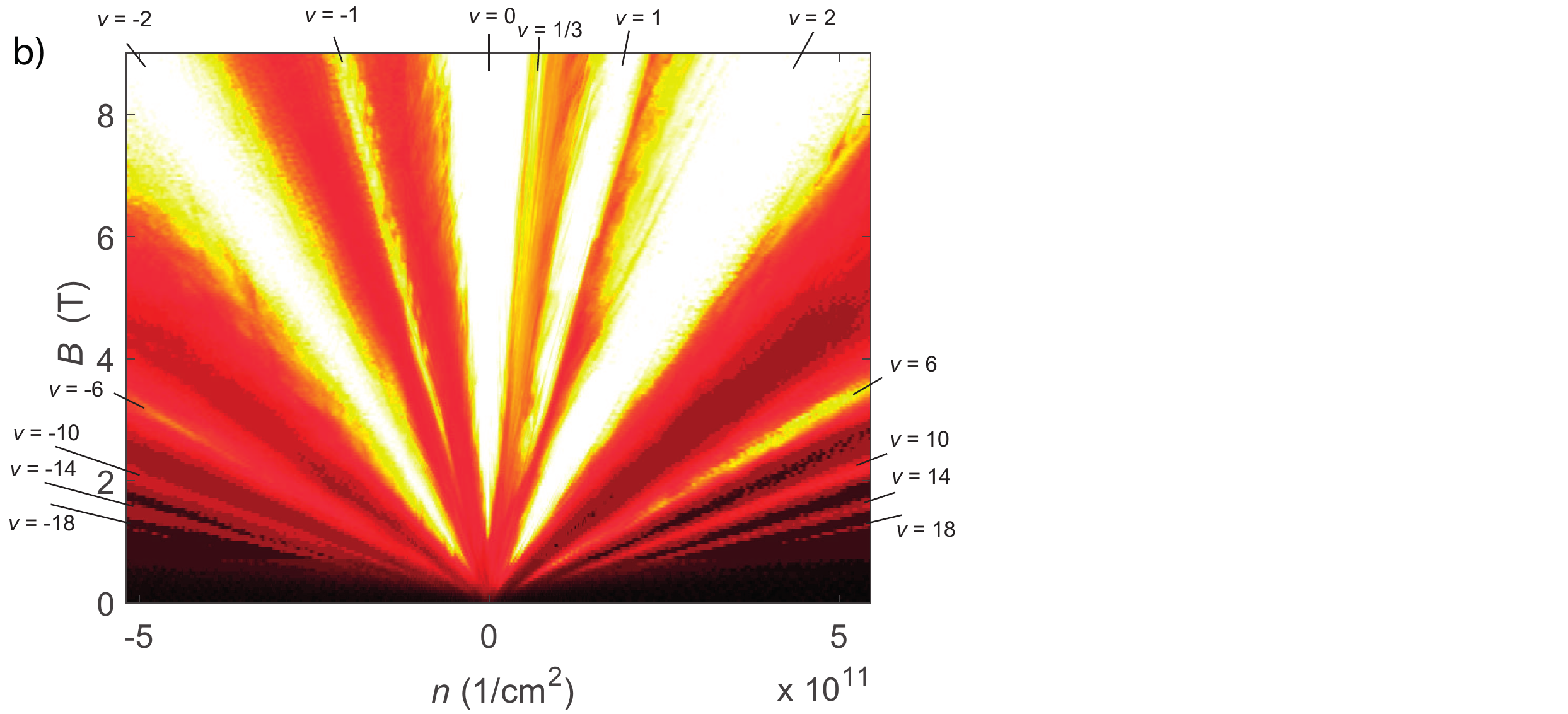}
\centering
\caption{a) Scanning electron micrograph of sample EV3 (on the left side); the scale is indicated by the white bar. b) A magneto-conductance Landau "fan plot" recorded on sample C2 on the plane spanned by charge density $n$ and magnetic field $B$. The charge density range corresponds to gate voltages $V_g \in$ [-60,60] V. The main filling factors are indicated in the picture.}
\label{fig:basics}
\end{figure}

Resists with different selectivities (lift-off-resist LOR for support and PMMA for lithography) were employed to facilitate our sample fabrication (see the supplemental material of Ref. \cite{Kumar2016}). We exfoliated graphene (Graphenium, NGS Naturgraphit GmbH) using a heat-assisted technique~\cite{Huang2015}. Monolayer graphene flakes were located on the basis of their optical contrast and the identification was verified using a Raman spectrometer with He-Ne laser (633 nm). The first contact defining the outer rim of the Corbino disk (see Fig. \ref{fig:basics}a)  was made in the regular manner \cite{Tombros2011}, and later the inner contact was fabricated together with a self-standing bridge to connect the inner electrode to a bonding pad. The strongly doped silicon Si++ substrate with a 285-nm layer of thermally grown SiO$_2$  provided the back gating electrode for the sample.

Initially after the fabrication process, our suspended devices tend to be $p$-doped in the first resistance $R$ \textit{vs.} gate voltage $V_g$ scans. Following the initial characterization, the samples were cooled down to $T$ = $10$ mK base temperature of the dilution refrigerator. Prior to DC and noise characterization, all devices were current annealed at the base temperature. These samples on LOR were typically annealed at a bias voltage of 1.6$\pm$0.1 V which is comparable with the optimal annealing voltage of our HF etched, rectangular two-lead samples \cite{Laitinen2014}. Subsequently, the samples were characterized by lock-in conductance measurement $G(V_g)$ to determine the mobility which was found to be $\mu > 10^5$ cm$^2$/Vs. The gate voltage was converted into charge carrier density by $n = (V_g-V_g^D)C_g/e$, where $V_g^D$ denotes the offset of the Dirac point from $V_g = 0$ V, typically these samples had $V_g^D \simeq -2$ V. A Landau fan plot, a conductance map on the $n-B$ plane, is displayed in Fig. \ref{fig:basics}b for sample C2. The gate capacitance  $C_g$ was obtained by fitting this Landau fan plot to the calculated locations  of the higher Landau levels  on the  $n-B$ plane.

The results of this paper cover two measured samples, C2 and EV3, with practically identical results. The inner and outer diameters of the EV3 Corbino ring were $d=0.8$ $\mu$m and $D=3.2$ $\mu$m, respectively. For the sample C2 the corresponding dimensions were $d=0.9$ $\mu$m and $D=2.8$ $\mu$m. The air gap between the graphene and the substrate surface, i.e. the LOR double layer thickness, was around 500 nm for both samples. The distance between  the back gate and graphene was compared with the value obtained from the gate capacitance by using the parallel plate capacitor model, which resulted in a reasonable agreement.

\subsection{Noise measurements}

The correlations in electron dynamics can be quantified using the Fano factor $F=S_I/S_I^P$ for the relative magnitude of current fluctuations \cite{Blanter2000}. The value $F = 1$ corresponds to Poissonian noise $S_I^P$, indicating non-correlated electron motion without any interactions between the electrons. For correlated motion, for example due to Pauli exclusion principle, the Fano factor becomes $F < 1$ and ultimately vanishes altogether for fully ballistic transport channels. Superpoissonian noise, on the other hand, is an indication for bunching of particles \cite{Blanter2000}, which can be observed e.g. in avalanche diodes \cite{Reulet2009}. In avalanche type of transport, charge carriers are grouped together through carrier multiplication. Such avalanche pulses can be considered at long time scales as a single charge carrier, which means that the resulting shot noise (assuming random triggering of the pulses) can be written as $S_I = F_{AV}2e\langle I \rangle$, where $F_{AV}$ is related to the average charge of the pulse $N_{tot}e = \tau_s I_0$ by $F_{AV} = 2N_{tot}$ according to Eq. \ref{limit1}.

On short time scales, the avalanche pulses do have fluctuations that originate from the specific excitation processes leading to the carrier multiplication. It is assumed in BSEH theories that thermal excitation is relevant in the carrier multiplication processes. These thermal processes are strongly temperature dependent, but they can last on the order of microseconds at the lowest temperatures (See, \textit{e.g.} Ref. \cite{Walsh2017}). Consequently, shot noise experiments around frequencies of 1 GHz are able to probe possible thermal-relaxation-induced correlations in the avalanche regime as the temperature due to the Joule heating of the sample increases with current.

Our low-frequency noise measurements were carried out using voltage bias up to $100$ mV. The current was recorded using a transimpedance amplifier (Stanford Research SR570, gain $10^6$ V/A) and its fluctuations were measured using a SRS 785 FFT analyzer. For the low-frequency noise power spectral density, we employed fast Fourier transform (FFT) with $800$ points spanning the range $125$ mHz - $100$ Hz, using a total measurement time of $150$ s per bias point. In our experiments, we are limited in the measurement bandwidth by the $RC$ cut-off, dependent on the resistance of the sample $R$ and the total capacitance of the lines $C\simeq 30$ nF.  Hence, in the results section, we characterize the bias dependence of the low-frequency noise at $10$ Hz which is the maximum frequency not yet appreciably influenced by the $RC$ cut-off frequency.

\begin{figure}[t]
\includegraphics[width=0.8\textwidth]{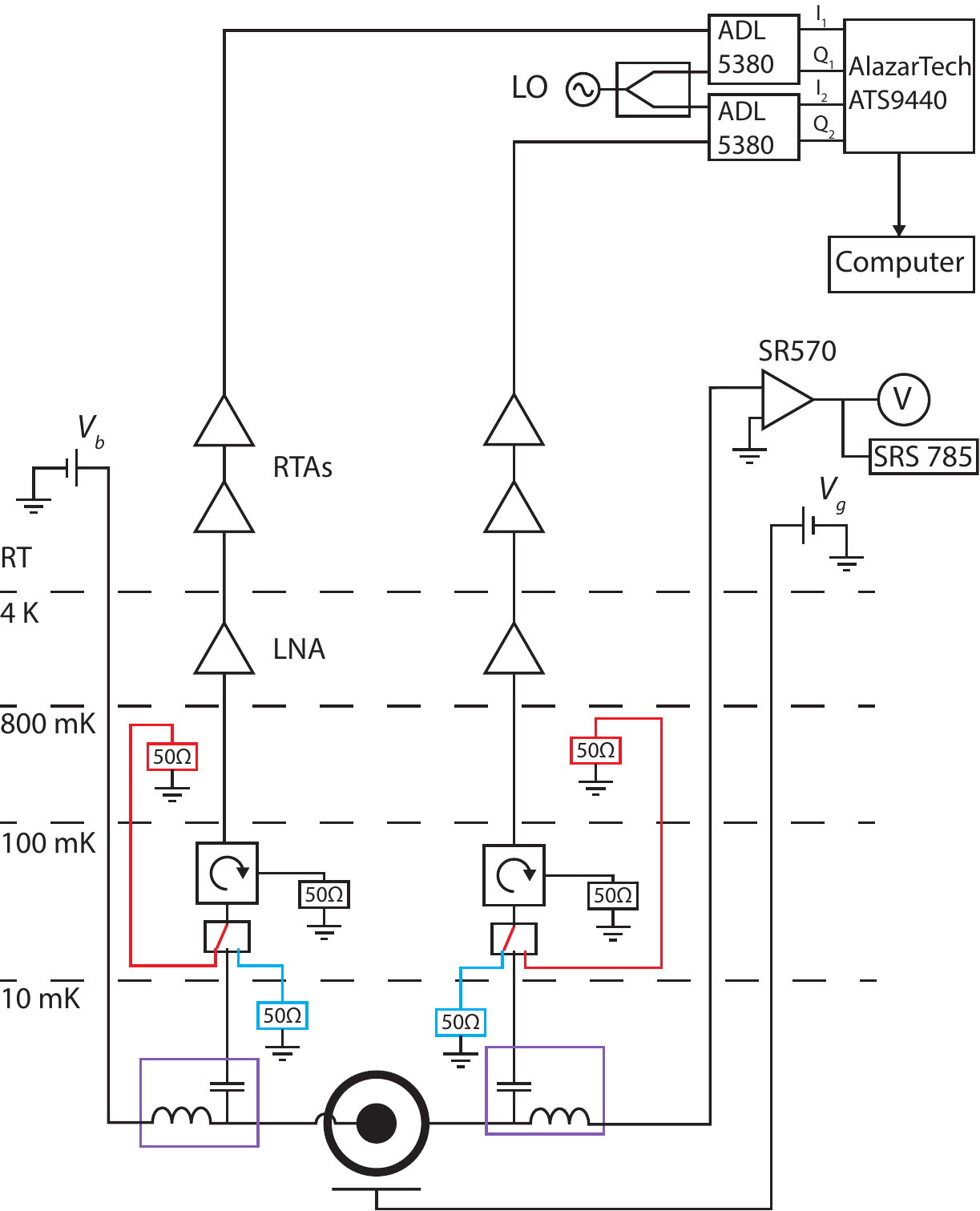}
\centering
\caption{Schematic of the measurement setup. The low- and high-frequency circuits are separated from each other by bias-Tees at low temperatures. The high-frequency side contains a microwave shot noise measurement hooked up to the contact electrodes of the Corbino ring, while the low frequency side is employed for DC-biasing as well as for measuring low-frequency noise from the outer contact.  There are separate high frequency coaxial cables with 50 $\Omega$ terminators acting as thermal noise sources, which can be coupled to the amplification chain through a microwave switch. The cooled low-noise amplifiers (LNA, Caltech CITLF3) and the room temperature $\mu$W-amplifiers (2 $\times$ Mini-Circuits ZRL-1150LN+) provide a gain of 100 dB in total.}
\label{fig:shotNoiseSchem}
\end{figure}

The high frequency noise was measured over frequencies $650 - 900$ MHz. Our microwave noise techniques follow the basic principles outlined in Refs. \cite{wu2006,wu2007,danneau2008,Nieminen2016}. In this work, we measured the excess shot noise $S_I(V) - S_I(0)$ using DC bias alone, while the zero-bias value $S_I(0)$ was measured  intermittently in the middle of each bias sweep for a drift correction. In order to avoid external spurious disturbances from mobile phones, the set-up was closed in a Faraday cage. The noise signal was first amplified by a cryogenic low-noise SiGe amplifier (Caltech CITLF3, gain 36 dB) with a nominal noise temperature of $T_{noise} \approx 4$ K. A circulator was used in both channels to block the amplifier noise reaching the sample, see Fig. \ref{fig:shotNoiseSchem}. The noise detection channel ends with two room-temperature amplifiers (Mini-Circuits ZRL-1150LN+, gain 32 dB) yielding the total gain of 92 dB, including an 8 dB attenuator used to limit the power to a suitable level. Finally, the signal was mixed down by a 780 MHz local oscillator and digitized using ADL-5380 IQ-mixers and a 125 MS/s AlazarTech ATS9440 digitizer card. The digitized data was auto- and cross correlated in real time using a computer in which GPU accelerated data processing was utilized. Custom software for the processing was written in CUDA C.

A microwave switch allowed us to  measure either shot noise from the graphene sample or thermal noise from $Z_0=50$ $\Omega$ resistors located at the mixing chamber (10 mK) and at the still plate (800 mK). These resistors allowed us to calculate the noise temperature of the measurement channels, and thereby calibrate the shot noise level. The system noise temperature of a measurement line amounts to:
\begin{equation}
T_{N} = \frac{T_{hot}-\frac{P_{hot}}{P_{cold}}T_{cold}}{\frac{P_{hot}}{P_{cold}}-1},
\end{equation}
where $P$ and $T$ refer to the thermal noise of a 50 $\Omega$ resistor and its physical temperature, while the labels $"hot/cold"$ refer to mixing chamber / still positions in the refrigerator, respectively.

One can use the system noise temperature $T_N$ to estimate the equivalent shot noise temperature from the excess noise correlation factor $C_{12}' = \frac{C_{12}}{\sqrt{C_1^2C_2^2}}$, where $C_i$ is the autocorrelation of channel $i$  and $C_{12}$ is the un-normalized cross-correlation between channels 1 and 2. $C_{12}'$ is  normalized using auto-correlations $C_i$ which are dominated by the system noise temperature ($2eIZ_0 \ll T_N$). Then, the noise temperature corresponding to the excess noise is given by:
\begin{equation}
T_e = C_{12}'T_N.
\end{equation}
Thus, the calibrated current  noise spectral density ($S_I$ couples fully to $Z_0$ due to the circulator) is obtained from $S_I = 4 k_B T_e/Z_{0}$, where $Z_0=50$ $\Omega$ is the characteristic impedance of the microwave system.

The validity of the calibration was tested by comparing the obtained small-bias Fano factor $F(V_g)$ at $B=0$ T to the theoretically predicted values \cite{Rycerz2009}. We reached an agreement within $\pm 20$ \%, but this may  reflect rather the insufficiency of the theoretical modelling than the inaccuracy of our calibration.

\section{Results}

\subsection{Low frequency switching noise}

\begin{figure}[htb]
\includegraphics[width=0.43\textwidth]{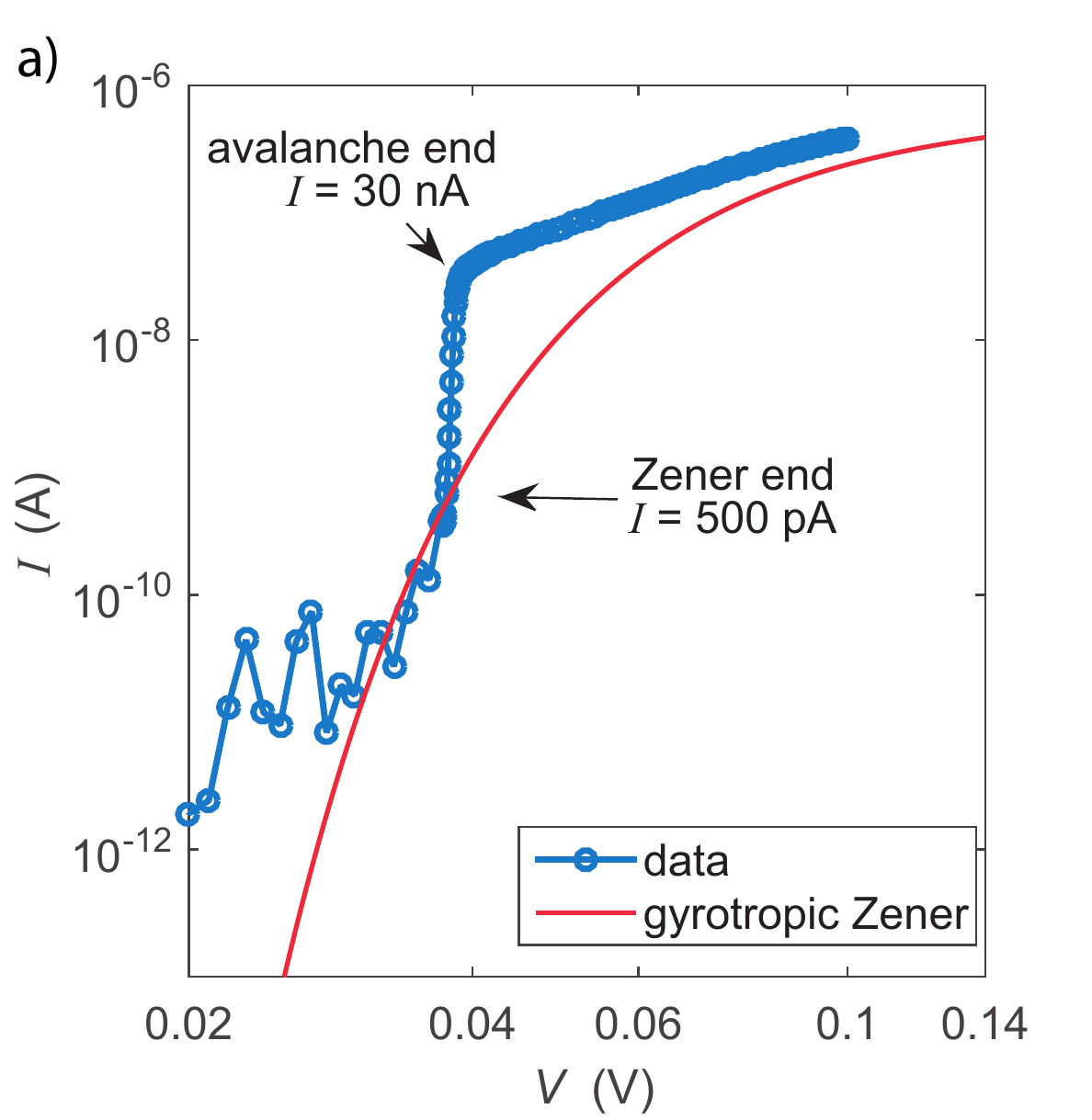}
\includegraphics[width=0.56\textwidth]{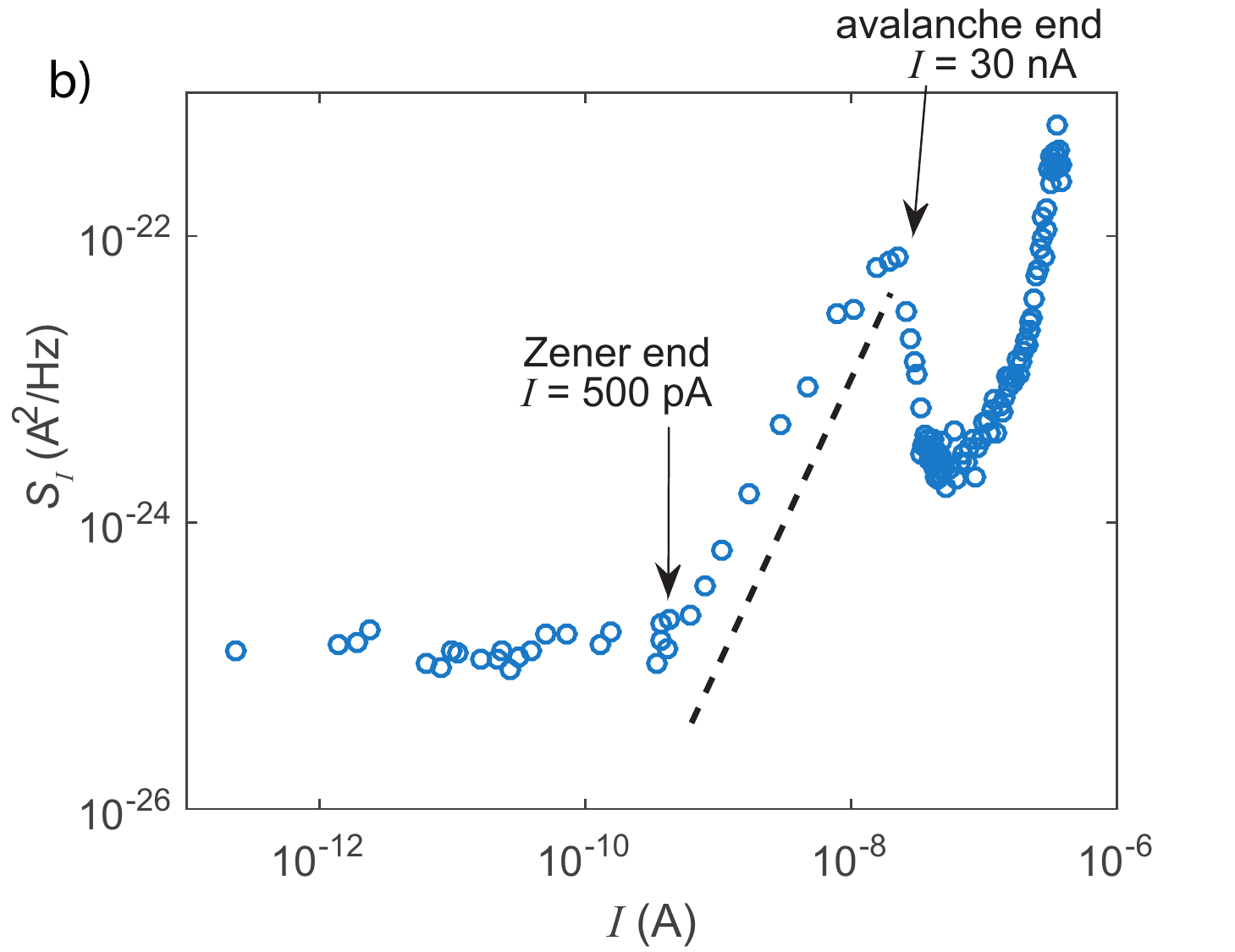}
\centering
\caption{a) IV characteristics measured on the sample EV3 at $B$ = 5.6 T and a fit to the data using the gyrotropic Zener tunneling model of Eq. \ref{vor}; positive bias was applied to the inner Corbino contact while the outer was connected to virtual ground. b) The low-frequency noise at $f$ = 10 Hz measured at the same time as the IV data. The black dashed line denotes the quadratic $I^2$ behavior.}
\label{fig:pinknoise}
\end{figure}

The $IV$ characteristics of sample EV3 measured at $B$ = 5.6 T is displayed in Fig. \ref{fig:pinknoise}. The data are compared with the theoretical model for the gyrotropic tunneling of Eq. \ref{vor}  illustrated by the red curve. There is an agreement between the model and the data only at small bias, above which  there is an abrupt increase in the current. Compared with the data at 2 T presented in Ref. \cite{Laitinen2017}, the current increase takes place at smaller current, and  the agreement with the gyrotropic Zener tunneling is over a smaller range. As already tentatively discussed in Ref. \cite{Laitinen2017}, the steep increase in current signifies the onset of the avalanche type of transport, which is verified in the low-frequency noise behavior. The beginning and the end of the avalanche regime are denoted by arrows at 500 pA and 30 nA, respectively.

The low-frequency noise recorded at 10 Hz is depicted in Fig. \ref{fig:pinknoise}b as a function of the bias current, extending across the avalanche type of transport towards ohmic behavior. The steep section of the IV curve between $0.5 - 30$ nA is seen to display noise that increases as $\langle I \rangle^2$ with bias current. This kind of current dependence is typical to switching noise as well as for resistance fluctuation noise~\cite{Kogan}.  Because of the sharp cut-off of the $\langle I \rangle^2$ dependence, we argue that the observed behavior can be assigned to avalanche transport which results in switching type of noise. This type of pulse sequence was illustrated in Fig. \ref{fig:switchingNoise}a in Sect. \ref{AVtheory}. By fitting the maximum value of the low-frequency noise $S_I^{\textrm{max}} \simeq 10^{-22}$ A$^2$/Hz at 10--20 nA to the model of Eq. \ref{limit1}, we obtain $\tau_0 \sim 100$ ns for the time separation between the pulses in the fully developed switching sequence (assuming $\omega \tau_s \ll 1$). Using the earlier definitions, we observe that the low-frequency Fano factor corresponds to bunching with $F_{AV} \approx 1.3 \times 10^4$ at the noise peak \footnote{Here one needs to remember that this approximation assumes white spectrum for the low-$f$ noise}.

Initially, only the size of the avalanche pulses grows with bias and produces an $\langle I \rangle^2$-dependent increase in the low-$f$ noise. However, towards the end of the avalanche regime at $I=30$ nA, the upwards transition rate $1/\tau_0$ starts also to change. This is seen as a decrease in the current noise by an order of magnitude, before the noise starts growing again as $\sim \langle I \rangle^2$.  The ten-fold increase in the upwards transition rate brings the value of  $\tau_0$ close to the value of $\tau_s$ (see below), which means gradual overlapping of individual avalanche pulses. The second increase in low-$f$ noise is assigned to the contacts, which are known to have low-frequency resistance fluctuations even in the case of the best suspended samples \cite{Kumar2015}.

\subsection{Shot noise at microwave frequencies}\label{muWaveNoise}

\begin{figure}[htb]
\includegraphics[width=0.44\textwidth]{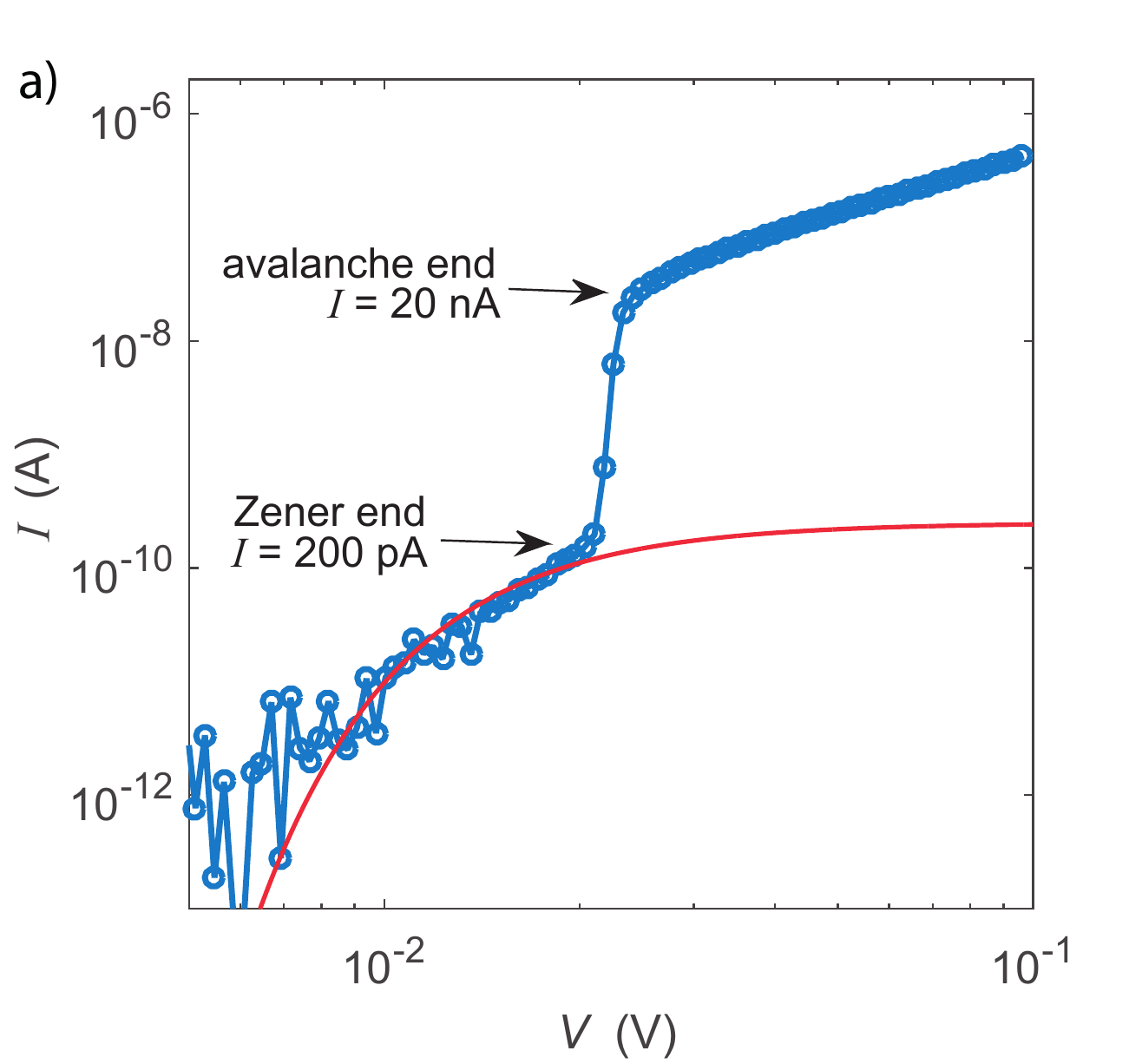}
\includegraphics[width=0.55\textwidth]{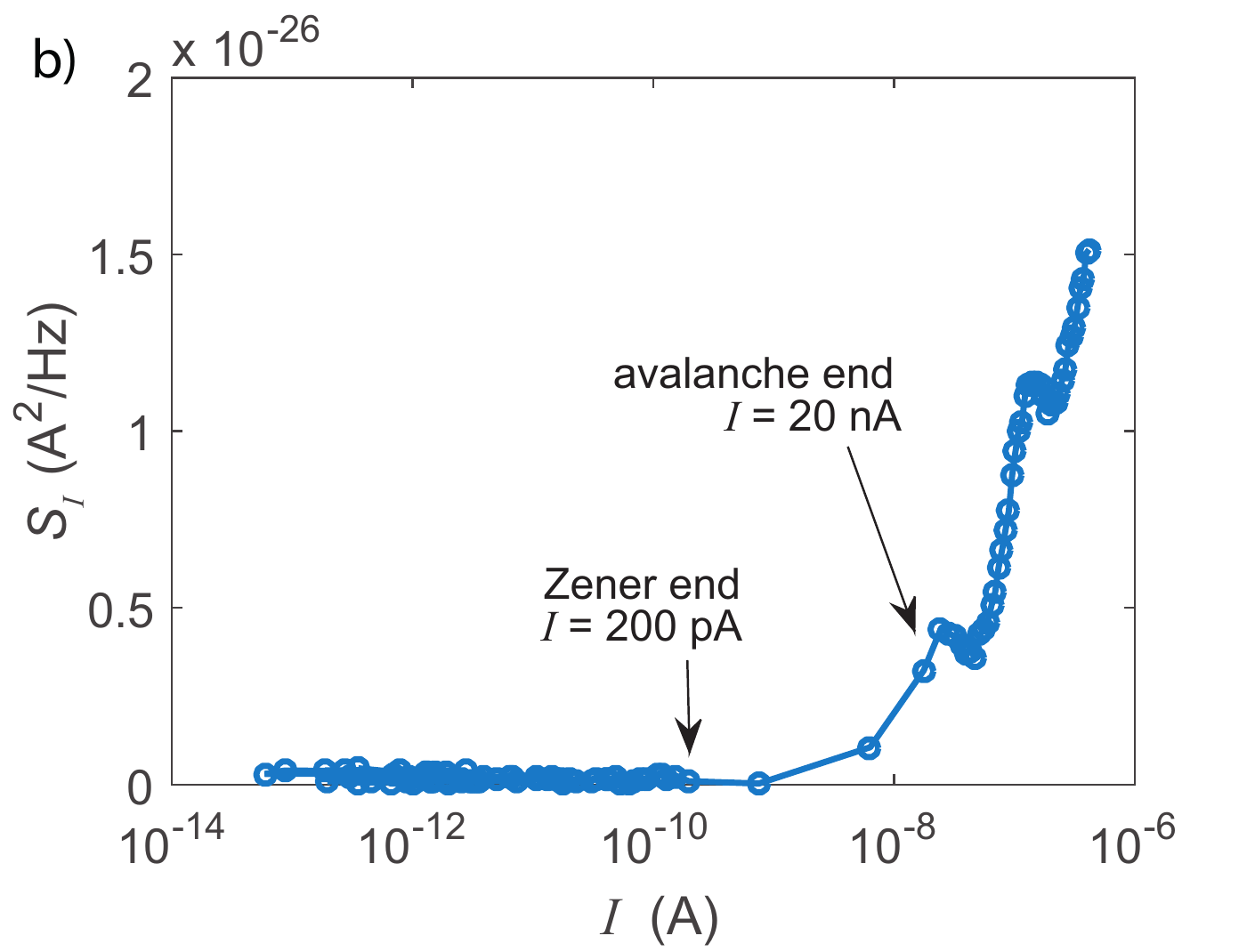}
\centering
\caption{a)  IV characteristics measured on the sample C2 at $B$ = 6 T and a fit to the data using the gyrotropic Zener tunneling model of Eq. \ref{vor}. The arrows indicate the beginning and the end of the avalanche transport regime. b) The shot noise spectral density (excess noise)  measured at microwave frequencies $f = 650-900$ MHz and recorded simultaneously with the IV data in the panel 5a. The arrows denote the same spots as in the IV picture; the increase of the shot noise across the avalanche regime is clearly seen.  }
\label{fig:shotnoise}
\end{figure}

To supplement the information on the low-frequency noise of the avalanche pulse, we have recorded shot noise at microwave frequencies $f=650-900$ MHz. The microwave shot noise data obtained on sample C2 at $B = 6$ T are illustrated in Fig. \ref{fig:shotnoise} together with the measured IV characteristics. The red curve in Fig. \ref{fig:shotnoise}a illustrates the behavior according to the gyrotropic Zener tunneling model at low bias. The arrows mark the beginning and the end of the avalanche transport regime at 200 pA and 20 nA, respectively.

Fig. \ref{fig:shotnoise}b displays the shot noise spectral density (excess shot noise $S_I(V) - S_I(0)$)  measured at microwave frequencies  recorded simultaneously with the IV data of Fig. \ref{fig:shotnoise}a; the arrows denote the same spots as in the IV picture. The shot noise displays a clear increase in the spectral density across the avalanche regime in which a large enhancement in $I$ and bunching of electrons takes place. At the end of the avalanche regime, the Fano factor $F \sim 1$ which points towards random Poissonian noise with uncorrelated charge carriers.  Above the avalanche regime, there is a decrease in the shot noise power which is similar, although weaker than was observed in the low-frequency noise. Thus, we conclude that when bunching decreases, as deduced from the effective low-$f$ noise Fano factor $F_{AV}$, the microwave shot noise also decreases. The variation in the shot noise power at microwave frequencies, however, remains much weaker than in the observations at low frequencies, and we find $F \lesssim 1$ for microwave excess noise in the avalanche regime, as well as above it.

\begin{figure}[htb]
\centering
\includegraphics[width=0.6\textwidth]{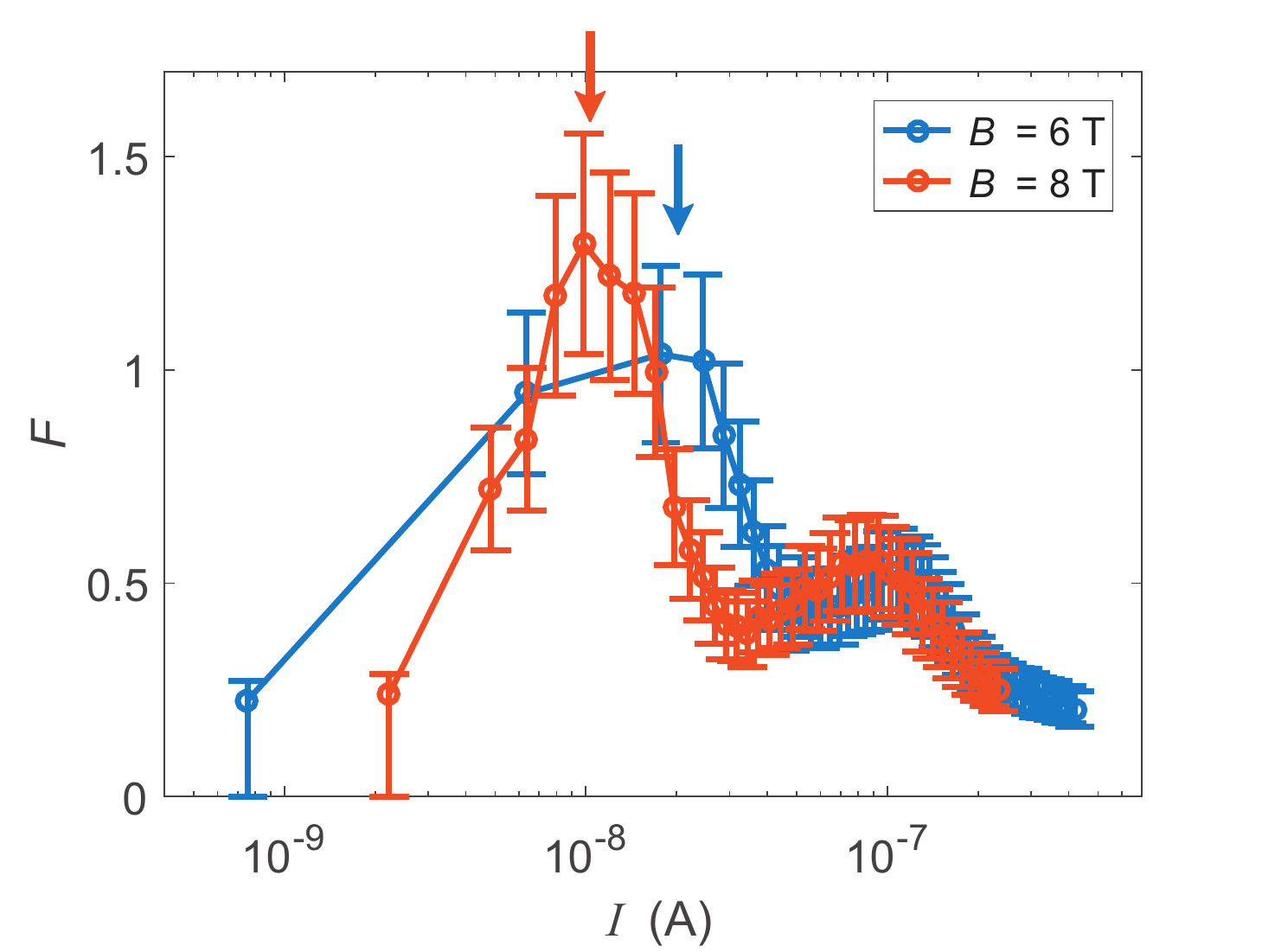}
\centering
\caption{The Fano factor as a function of the bias current $I$ for sample C2 at $B = $ 6 T and 8 T, obtained from the results such as in Fig. \ref{fig:shotnoise}b. The displayed data sets start from the beginning of the avalanche region, below which the noise measurement displays just the background noise of the measurement setup and $F$ cannot be defined. The end of the avalanche region is marked by the arrows for both data sets. Note the increase of the Fano factor within the avalanche regime.}
\label{fig:shotnoise2}
\end{figure}

The Fano factors for magnetic fields $B=6$ and 8 T as deduced from the  noise power spectral density are displayed in Fig. \ref{fig:shotnoise2}a and b, respectively. Both data sets indicate an increase of the Fano factor when going deeper into the avalanche regime. Initially, just above the avalanche threshold, the Fano factor is small, $F \sim 0.2$ as expected for a time-correlated sequence of  electrons within the avalanche pulse.  The increase of $F \rightarrow 1$ within the avalanche regime is in line with a development of a single multiplication site to multiple ones, which would lead to reduced temporal correlations between electrons in the generated charge pulse, and thereby to an enhanced Fano factor; the observed maximum value $F = 1.2 \pm 0.2$, however, would suggest partly simultaneous triggering events of the multisite generation. Above the avalanche regime, the Fano factor clearly decreases to a level $F \simeq 0.5$ at bias voltages $\sim 40$ mV. This plateau is close to what one would expect for  hot-electron  transport in a diffusive conductor, but such theories exist only for the half-filled 2-DEG Landau level case, \textit{i.e.} for a composite fermion Fermi sea \cite{vonOppen1997}. At high bias, the Fano factor starts to diminish again, which indicates the presence of inelastic processes, such as the electron - phonon coupling, leading to a suppression of the noise. Our data display $F\simeq0.2$ at 100 mV and a power-law-like decrease as $V^{-1} \ldots V^{-2}$.

In the BSEH model, heating of the electron gas plays a central role. The Joule heating of the electron gas has to be transported away either via phonons or via electronic thermal conduction. In the quantum Hall regime, the electronic conductance is weak, which makes the electronic thermal conductivity negligible in our breakdown experiments.  Therefore, the thermal balance is governed by electron - phonon coupling. In our suspended graphene, the main cooling channel at small energies is via acoustic phonons while at large energies supercollisions with flexural phonons are the dominant process \cite{Laitinen2014}.  In our present experiment, the electron - phonon coupling will increase with field as the density of states of electrons increases linearly with the magnetic field, assuming that the width of the Landau level remains fixed. However, there is an opposite tendency arising from the shrinking of the electron wave function with $B$, which makes it difficult to couple the long-wave-length acoustic phonons to the high-field electrons and a reduced increase in the coupling results \footnote{In regular 2-DEG heterostructure, an increase in the electron - phonon coupling by a factor of two is found between 2 and 9 T. \protect{\cite{Prasad1984}}}.

\begin{figure}[tb]
\centering
\includegraphics[width=0.49\textwidth]{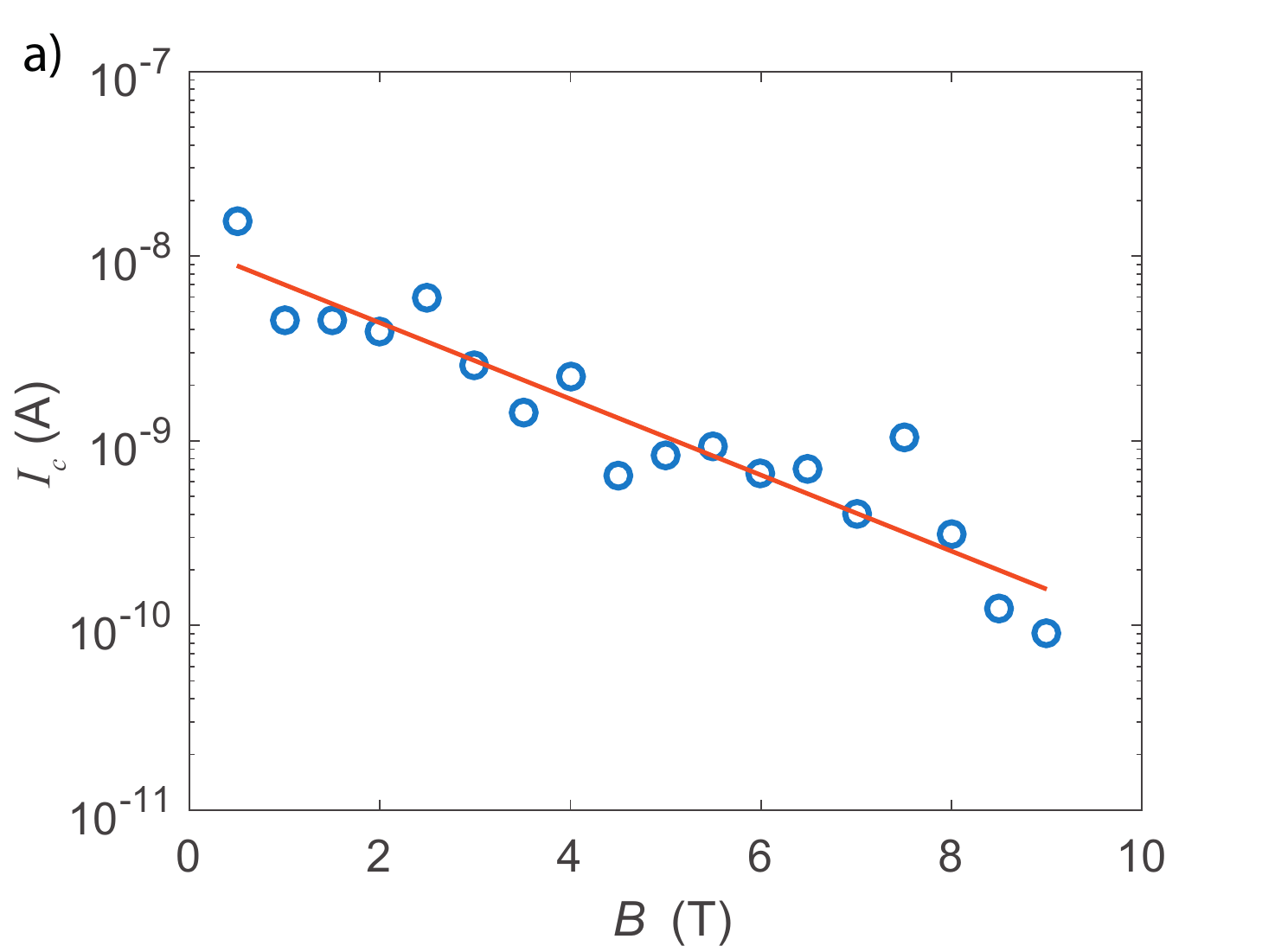}
\includegraphics[width=0.49\textwidth]{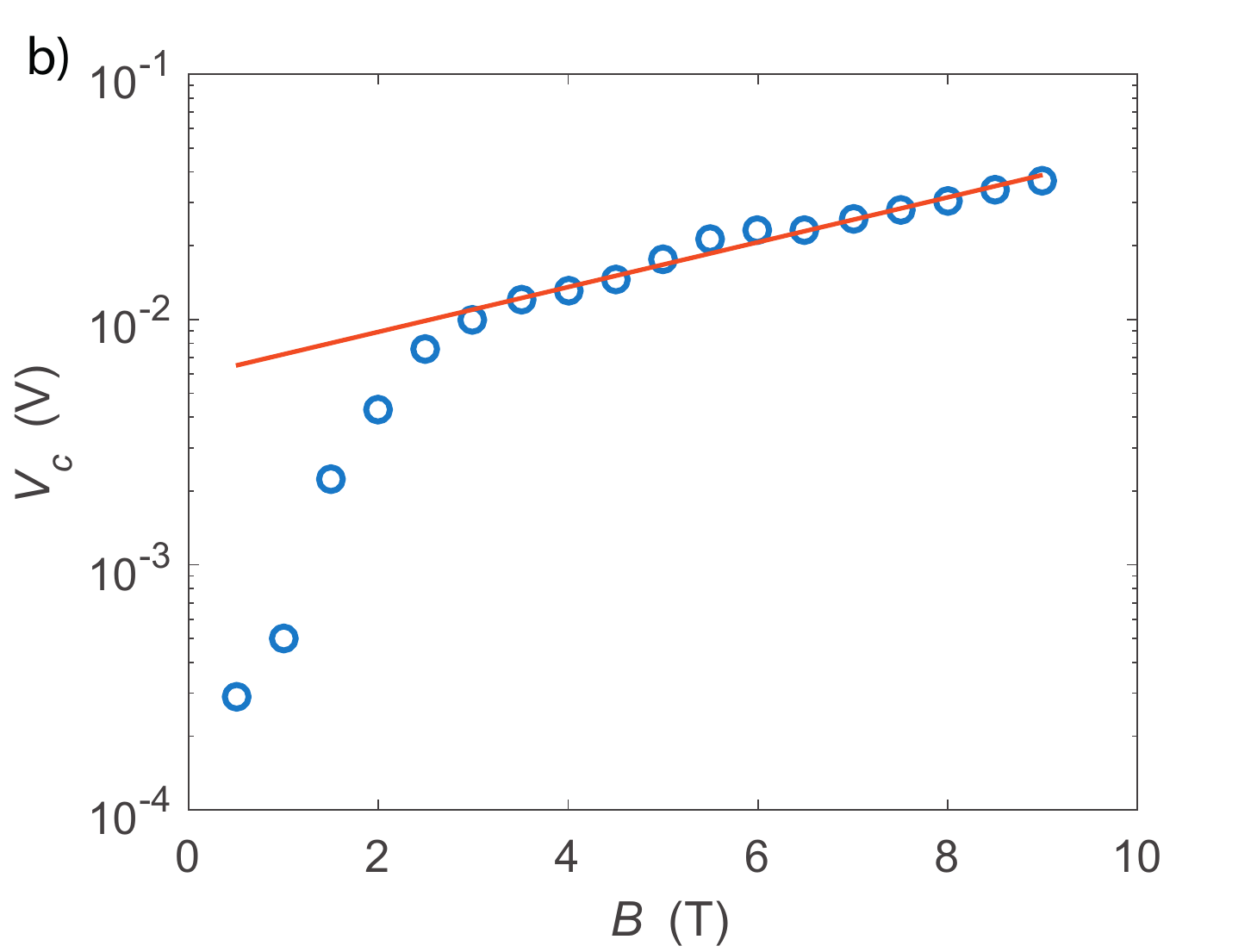}
\centering
\caption{Take-off point of the avalanche region, i.e. the point where the low-frequency noise begins to grow: a) the critical current $I_c$ as a function of magnetic field $B$, b) the corresponding critical voltage $V_c(B)$. Note that the data in Fig. \ref{fig:shotnoise2} deviate slightly from the data here; the data sets derive from different measurements. The deviation does not influence the conclusions on $I_c$ and $V_c$ at magnetic fields $B > 2$ T, which are based on the fitted solid lines given by $I_c=I_0 \times \exp{(-B/B_I)}$ and $V_c=V_0 \times \exp{(B/B_V)}$, where $I_0=1.1 \times 10^{-8}\textrm{ A}$, $V_0=5.8 \textrm{ mV}$, $B_I=2.1 \textrm{ T}$ and $B_V=4.8\textrm{ T}$. The product of these fits indicates that $I_c V_c^2$ is approximately constant at large magnetic fields.}
\label{fig:AvalancheBeginning}
\end{figure}

We have investigated the power required to initiate the avalanche type of transport by measuring both the critical current $I_c$ and the critical voltage $V_c$ for breaking the gyrotropic tunneling regime. In our data at fields $B > 2$ T displayed in Fig. \ref{fig:AvalancheBeginning}, we find  a decrease in the heating power $P=I_c V_c$, \textit{ i.e.} a smaller power $P$ is required to initiate the avalanche regime.  In fact, the quantity $ I_c V_c^2$ appears to be independent of the magnetic field. According to our shot noise results around the Dirac point, the electron-phonon coupling is approximately independent of the magnetic field at $B > 2$ T. These results together suggest that if the BSEH processes become active at some fixed temperature, then the heating power needs to be deposited to an area that decreases with the bias voltage. One possibility is that the power is dissipated into a region that extends between the two tilted Landau level bands: their spatial separation decreases with increasing bias.

\section{Discussion}\label{disc}

To our knowledge, our results are the first ones dealing in detail with the crossover from  Zener tunneling of single electrons to BSEH behavior leading to avalanche type of transport. The crossover is easily visible at moderately low magnetic fields ($\sim 2$ T), but it becomes exceedingly difficult to distinguish the gyrotropic Zener tunneling regime at magnetic fields above 8 T. The maximum observable Zener tunneling current decreases by two orders of magnitude between 1 and 9 Tesla in our experiments. This indicates that avalanche type of breakdown of the zero-energy Landau level becomes very easily triggered at high magnetic fields.

Our low-frequency noise results in the avalanche regime are similar to those of Kobayashi and coworkers on GaAs Corbino rings~\cite{Chida2014,Hata2016}. In both experiments, very strong bunching of charge carriers is observed, which supports the view of BSEH type of carrier excitation. We note that our sample size is approximately by a factor of 50 smaller than the GaAs devices investigated in Refs. \cite{Chida2014,Hata2016}, and yet, the carrier excitation seems to be nearly equally efficient as judged from our observed avalanche-regime Fano factors $F_{AV} = 1.3 \times 10^4$ at 5.6 T, in comparison to $F_{AV} = 10^3 - 10^5$ found in Refs. \cite{Chida2014,Hata2016}.

In addition to the low-frequency noise, we also probed the noise at microwave frequencies and found predominantly sub-Poissonian shot noise in this case, with a Fano factor varying in the range $F = 0.2 - 1.2$ at $B =$ 8 T. The Fano factor at microwave frequencies measured at the end of the avalanche regime seems to be quite independent of the magnetic field. The value of the Fano factor suggests that the Lorentzian spectrum caused by the switching noise has to decay below the shot noise before the GHz frequency range. If we take $ F_{AV}=10^4$, then $\omega \tau_s > 100$ is a necessary condition, which means that the avalanche pulse duration has to satisfy $\tau_s > 20$ ns. As this pulse contains $5 \times 10^3$ electrons, the average time distance of the electrons becomes $\geq 4$ ps, which corresponds to only $\leq 50$ nA in average current during the avalanche pulse. Note that this value is not far from the current value at the end of the avalanche regime, which would then just correspond to the beginning of the overlap of the avalanche pulses.

In the light of the rather weak conductance during the avalanche pulses, the observed values of $F \simeq 1$ at the end of the avalanche regime may signal that the charge carrier emission events within the avalanche pulse become quite random at large currents. This could indicate several parallel transport paths where random carrier escape is supported by elevated local temperature. Such a state would transform smoothly to ohmic behavior with increased Joule heating by the bias.

One possible theoretical framework for a single transport path at high bias is provided by transport in a 1-dimensional array of tunneling junctions \cite{Golubev2004}. In this case, the Fano factor is around $F=1/3$ but it may vary substantially depending on the properties of each scattering/tunneling element (\textit{i.e.} their $F$ and $R$). Furthermore, array models can be extended to two dimensions where solitons may produce avalanche-like behavior and a suitable increase of the Fano factor around Coulomb blockade energies $E_c$, matching with our finding. \cite{Sverdlov2001}. The increase in Coulomb blockade voltage $E_c/e$ as $1/\ell_B$ is also in agreement with our observed upward voltage shift in the microwave Fano factor $F(V)$ with magnetic field.

Since the theoretical models are able to give reasonable explanations for $F \gtrsim 0.3$ in the coherent transport regime, we conclude from our results that strong correlations between the electrons within an avalanche pulse of electrons exist only near the onset of the avalanche regime (where $F \lesssim 0.23$). The number of electrons in such a pulse is $N_{\textrm{tot}} \simeq 10^3$ according to the low-frequency noise measurements.

\section{Conclusions}

The low-frequency noise clearly distinguishes between the gyrotropic Zener tunneling regime and the avalanche type of transport in the 0$^{th}$ Landau level of graphene. With an increasing magnetic field, the avalanche type of behavior becomes more favorable, and above $B =$ 8 Tesla it is hard for us to distinguish any Zener tunneling regime. The low-frequency noise in the avalanche regime displays features which are distinct to switching noise, i.e., the noise grows quadratically with the bias current. At the largest noise levels, this noise equals to an effective Fano factor on the order of $F_{AV} \simeq 10^4$, which also yields  an estimate of 10 MHz for the switching rate of the avalanche pulses with a duration of $> 20$ ns. Our measurements of the high-frequency microwave shot noise indicate clear correlations within the one-thousand-electron avalanche pulses at the onset of the avalanche transport. However, we also find that these charge carriers within the avalanche pulses become less and less correlated with increasing bias. This is seen as growth of the microwave $F$ across the avalanche region, at the end of which we obtain $F = 1.2 \pm 0.2$. In the high bias transport regime, the Fano factor is lowered as $V^{-1} \ldots V^{-2}$ and amounts to $F \simeq 0.2$ at 100 mV, which is in line with inelastic processes caused by electron - phonon interactions.

\begin{acknowledgements}
We thank C. Flindt, A. Harju, T. Ojanen, S. Paraoanu, and B. Pla\c{c}ais, for fruitful discussions. This work has been supported in part by the EU Framework Programme (FP7 and H2020 Graphene Flagship), by ERC (grant no. 670743), and by the Academy of Finland (project no. 250280 LTQ CoE). A.L. is grateful to Vais{\"a}l{\"a} Foundation of the Finnish Academy of Science and Letters for a scholarship. This research project made use of the Aalto University OtaNano/LTL infrastructure which is part of European Microkelvin Platform.
\end{acknowledgements}


%
%

\end{document}